\documentclass[twocolumn,secnumarabic,amssymb, nobibnotes, aps, prl, superscriptaddress]{revtex4-1}

\usepackage{float}
\usepackage{graphicx}
\usepackage{subfigure}
\usepackage{amsmath} 
\usepackage{epstopdf}
\usepackage{setspace}
\usepackage[shortlabels]{enumitem}
\usepackage{mathrsfs}
\usepackage{subfig}
\usepackage{color,soul}
\usepackage[dvipsnames]{xcolor}
\usepackage{dcolumn,amssymb,subfig}
\usepackage{bm}
\usepackage{ulem}

\renewcommand{\vec}[1]{\boldsymbol{#1}}
\newcommand{\tens}[1]{\boldsymbol{#1}}
\newcommand{\bnabla}{\vec{\nabla}}

\begin{document}
\title{Active nematics with anisotropic friction: the decisive role of the flow aligning parameter}

\author{Kristian Thijssen}
\affiliation{The Rudolf Peierls Centre for Theoretical Physics, Department of Physics, University of Oxford, Parks Road, Oxford OX1 3PU, UK}
\author{Luuk Metselaar}
\affiliation{The Rudolf Peierls Centre for Theoretical Physics, Department of Physics, University of Oxford, Parks Road, Oxford OX1 3PU, UK}
\author{Julia Yeomans}
\email{Julia.Yeomans@physics.ox.ac.uk}
\affiliation{The Rudolf Peierls Centre for Theoretical Physics, Department of Physics, University of Oxford, Parks Road, Oxford OX1 3PU, UK}
\author{Amin Doostmohammadi}
\email{doostmohammadi@nbi.ku.dk}
\affiliation{The Niels Bohr Institute, University of Copenhagen, Blegdamsvej 17, 2100 Copenhagen, DK}

\begin{abstract}
We use continuum simulations to study the impact of anisotropic hydrodynamic friction on the emergent flows of active nematics. We show that, depending on whether the active particles align with or tumble in their collectively self-induced flows, anisotropic friction can result in markedly different patterns of motion. In a flow-aligning regime and at high anisotropic friction, the otherwise chaotic flows are streamlined into flow lanes with alternating directions, reproducing the experimental laning state that has been obtained by interfacing microtubule-motor protein mixtures with smectic liquid crystals. Within a flow-tumbling regime, however, we find that no such laning state is possible. Instead, the synergistic effects of friction anisotropy and flow tumbling can lead to the emergence of bound pairs of topological defects that align at an angle to the easy flow direction and navigate together throughout the domain. 
In addition to confirming the mechanism behind the laning states observed in experiments, our findings emphasise the role of the flow aligning parameter in the dynamics of active nematics. 
\end{abstract}

\maketitle


\section{Introduction}\label{intro}

Active matter describes systems such as bacterial biofilms or cellular tissues that extract energy from their surroundings at the single-particle level and convert it into mechanical work manifest in the form of self-propulsion and active stress generation~\cite{Ramaswamy10, Marchetti13, Prost2015}. The continuous injection of energy - or activity - can lead to phenomena such as collective motion~\cite{Dombrowski2004,Sanchez2012,Sumino2012}, and active turbulence \cite{Wensink2012,Dunkel2013}. In addition to their intrinsic activity, the dynamics of active materials can be influenced by the shape of their constituent particles. In particular, elongation of particles is an important factor in determining their alignment dynamics and the emergence of orientational order in active materials. Indeed, an increasing number of biological active systems including colonies of rod-shaped bacteria~\cite{Volfson08,DoostmohammadiPRL2016,Li19}, cellular monolayers~\cite{Duclos2017,Saw2017,Kawaguchi2017}, and subcellular filaments~\cite{Sanchez2012,Keber2014,Guillamat16,Kumar18}, display orientational order and topological defects, singular points in the orientation field,  that resemble features of nematic liquid crystals. These materials are commonly described as active nematics~\cite{Ramaswamy10, Marchetti13,Doostmohammadi2018}.

Linear stability analysis shows that in infinite systems a two-dimensional active nematic is unstable to any level of activity in unconfined frictionless systems~\cite{Simha2002,Ramaswamy2007}. Numerical simulations have further shown that as this instability grows, there is a transition to active turbulence, a state characterised by chaotic flows and motile topological defects \cite{Giomi2013,Thampi2013}. 
This immediately points to the challenge of constraining and directing active materials.

Recent studies have, however, shown that inducing hydrodynamic screening can stabilise the chaotic flow patterns of active matter~\cite{Bechinger2016,Doostmohammadi2019}. One way of introducing the screening is by confining active matter into circular or rectangular geometries~\cite{Voituriez2005,Wensink2012,Doostmohammadi2017}. This can result in oscillations~\cite{Notbohm2016,Pearce2019,Shendruk2017,Opathalage2019,Hardouin2019,Felix2019} or unidirectional motion~\cite{Wioland2013,Doxzen2013,Wioland2016,Wu2017}. It is also possible to introduce hydrodynamic screening through frictional damping between active materials and their surroundings. Linear stability analysis shows that frictional damping increases the activity threshold required for the emergence of hydrodynamic instabilities in unconfined systems \cite{Thampi2014PRE}. Furthermore, numerical studies show that such frictional damping can stabilise active nematics into vortex-lattices~\cite{Doostmohammadi2016} and control alignment of topological defects~\cite{Pearce2019friction}. 
Of particular interest, recent experiments have shown that interfacing microtubule-motor protein mixtures with anisotropic surfaces of smectic liquid crystals can streamline microtubules into a `laning' state of jets flowing in alternating directions parallel to the smectic layers. This indicates that introducing anisotropy in the hydrodynamic screening can work as a potential mechanism for directing active flows into predesigned directions.

In this paper, we numerically investigate the impact of anisotropic hydrodynamic screening by subjecting a two-dimensional active nematic to anisotropic frictional damping. By systematically varying the strength of the anisotropy, we obtain the laning state observed  in the experiments on microtubule-kinesin motor mixtures. Interestingly, we find that this state is only possible in the flow-aligning regime and we explain the mechanism for the emergence of lanes in terms of the combined effects of frictional anisotropy and the alignment of active particles with the extensional velocity gradients. 
 Moreover, by changing the flow aligning parameter to values  deep within the flow-tumbling regime, we observe significant changes in the alignment of comet-like ($+1/2$) topological defects and find an unexpected state of bound defect pairs that navigate through the system  leaving long-lived distortions within the active nematic.

\section{Simulation method}
We employ the active nematohydrodynamic approach that is widely used to describe dense active nematics such as microtubule-motor protein mixtures and bacterial colonies~\cite{Giomi2012,Marchetti13,Doostmohammadi2018}. The dynamics of the system are described by the evolution of the incompressible fluid velocity $\vec{u}$ and the 
 orientational order parameter, the nematic tensor $\tens{Q}$:
\begin{eqnarray}
\bnabla\cdot\bm{u} & = & 0, \label{eq:u0} \\
\rho\left(\partial_t + \bm{u} \cdot \bm{\nabla}\right)\bm{u} & = & -\bnabla p + \bm{\nabla} \cdot \bm{\Sigma} - \bm{f} \bm{u}, \label{eq:momentum} \\
\left(\partial_t + \bm{u} \cdot \bm{\nabla}\right)\bm{Q} - \bm{S} & = & \Gamma_Q \bm{H} \label{eq:Q}
\end{eqnarray}
where $p$ is the pressure, $\rho$ is the density, $\tens{\Sigma}$ is the stress tensor, and $\bm{f}$ is a diagonal tensor describing the friction between the active nematic layer and the surroundings. In our model, this friction coefficient will depend on the direction of the velocity, corresponding to situations where the underlying substrate or the surrounding medium extracts momentum from the active layer in an anisotropic manner. The nematic tensor $\tens{Q}$ is defined as $\tens{Q}=\frac{3q}{2}\left(\bm{n}\bm{n}-\frac{1}{3}\bm{I} \right)$, where $q$ is the magnitude of the nematic order parameter and $\bm{n}$ is the headless direction of orientation.  

The co-rotation term $\bm{S} = \left(\lambda \bm{D}+\bm{\Omega}\right)\left(\bm{Q}+\frac{1}{3}\bm{I}\right) + \left(\bm{Q}+\frac{1}{3}\bm{I}\right)\left(\lambda \bm{D}-\bm{\Omega}\right)-2\lambda\left(\bm{Q}+\frac{1}{3}\bm{I}\right)\text{tr}\left(\bm{Q}\bm{W}\right)$ determines the alignment of elongated particles in response to gradients in the velocity field. $\tens{\Omega}$ is the rotational part of the velocity gradient tensor, $\tens{D}$ is the extensional part, and $\tens{W}$ is the total velocity gradient tensor. The alignment of particles with respect to the extensional and rotational components of velocity gradients is characterised by the flow aligning parameter $\lambda$ that is proportional to the aspect ratio of the particles. For $\lambda > 9q/(3q+4)$ the director aligns at a given angle to a shear flow (the Leslie angle), while for $\lambda \le 9q/(3q+4)$ 
the director field rotates (tumbles) in a simple shear~\cite{BerisBook}. The value of the flow-aligning parameter depends on the size, aspect ratio, magnitude of the order, and also interactions between the nematogens. 

Indeed, one of the few attempts to extract the flow-aligning parameter, for the wing epithelium of {\it Drosophila}, has shown that it can have a range of values and even become negative~\cite{Aigouy10}. Therefore it would not be surprising if different experimental systems show distinct behaviours associated with flow-tumbling or flow-aligning behaviour. However, to our knowledge, the role of this parameter has been only marginally explored in most theoretical works and experiments with active nematics.

The relaxational dynamics of the nematic tensor $\tens{Q}$ is governed by the molecular field $\bm{H} = -\left(\frac{\delta\mathcal{F}}{\delta\bm{Q}} - \frac{\bm{I}}{3}\text{tr}\frac{\delta\mathcal{F}}{\delta\bm{Q}}\right)$, and the rotational diffusion coefficient $\Gamma_Q$ sets the time scale for the relaxation.
The associated free energy $\mathcal{F}$ is composed of a bulk contribution
\begin{equation}
\mathcal{F}_\text{bulk} = A_0\left(\frac{1}{2}\Big(1-\frac{\gamma}{3}\Big)\text{tr}(\bm{Q}^2) - \frac{\gamma}{3}\text{tr}(\bm{Q}^3) + \frac{\gamma}{4}\text{tr}(\bm{Q}^2)^2\right)
\end{equation}
giving an isotropic-nematic transition at $\gamma=2.7$~\cite{Matsuyama2002}, and a Frank free energy term $\mathcal{F}_\text{el} = \frac{K}{2}\left(\bm{\nabla Q}\right)^2$ that penalises deformations in the orientation field. 

Using this free energy description, the passive part of the stress tensor can be written as
\begin{equation}
\begin{split}
\bm{\Sigma}^\text{passive} = 2\nu \bm{D} -\lambda\bm{H}\left(\bm{Q}+\frac{1}{3}\bm{I}\right)\\
-\lambda\left(\bm{Q}+\frac{1}{3}\bm{I}\right)\bm{H}+2\lambda\left(\bm{Q}+\frac{1}{3}\bm{I}\right)\text{tr}\left(\bm{QH}\right)\\
+\bm{Q}\bm{H}-\bm{H}\bm{Q}- \bm{\nabla Q}\frac{\delta \mathcal{F}}{\delta\bm{\nabla Q}},
\label{eq:sigma}
\end{split}
\end{equation}
where $\nu$ is the viscosity of the fluid. In addition to the passive part, the active stress is defined as $\bm{\Sigma}^\text{active} = -\zeta \bm{Q}$ such that gradients in the orientational order $\tens{Q}$ generate stresses that drive active flows, with $\zeta$ setting the strength of the activity.

Equations~(\ref{eq:u0})--(\ref{eq:Q}) are solved using a hybrid lattice-Boltzmann method~\cite{Marenduzzo2007}. Unless otherwise specified, the numerical parameters that we use are $\nu=2/3$, $\Gamma_Q=0.7$ and $p=0.25$. We use the single elastic constant approximation with $K=0.03$ and take the bulk free energy parameters as $A_0=1$ and $\gamma=2.85$, resulting in an equilibrium magnitude of the nematic order of $q_{eq}=0.24$. The default values of the parameters that we vary in this manuscript are the activity $\zeta=0.03$ and flow aligning parameter which is typically chosen as $\lambda=0.7$ and $\lambda=0.3$ which are respectively deep in the flow-aligning and flow-tumbling regime. These parameters are chosen in the range that reproduces different flow patterns of microtubule-kinesin motor mixtures in confinement~\cite{Shendruk2017}. Lastly, domain sizes are chosen as $250\times250$ lattice sites unless stated otherwise in the figure caption and all
simulations are performed with periodic boundary conditions.

\begin{figure*} 
    \centering
    \includegraphics[width=0.95\textwidth]{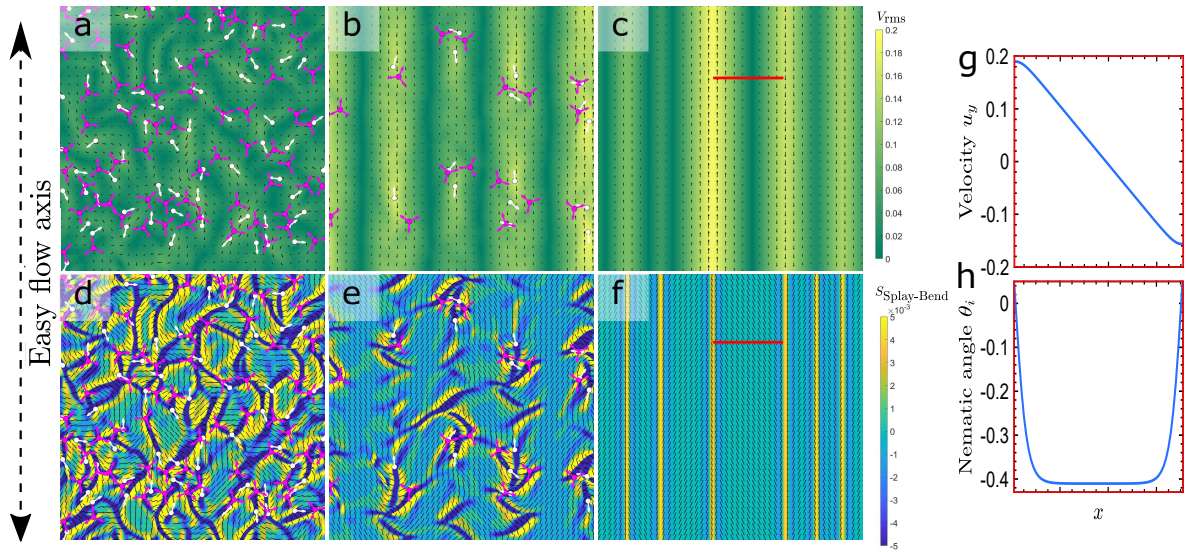}
    \caption{\textbf{Emergence of laning with increasing friction in the flow-aligning regime.}
Friction is applied along the $x$-axis making the $y$-axis the easy flow axis for the flow. 
(a)-(c) Velocity field coloured by the magnitude of the rms-velocity  $u_\text{rms}$. (d)-(f) Director field coloured by the magnitude of the splay-bend order parameter $S_\text{Splay-Bend} = \partial_i \partial_j Q_{ij}$. White circles denote $+1/2$ defects with the attached lines indicating their orientation (pointed from head-to-tail) and purple trefoils denote $-1/2$ defects. (a),(d): Low friction $f_x = 0.004$, showing active turbulence. (b),(e): Intermediate friction $f_x=0.015$, leading to a laning state with active instabilities. (c),(f): High friction $f_x=0.07$, showing a stable laning state. The activity in (a)-(f) is kept constant at $\zeta = 0.03$. (g) Variation of the $y$-component of velocity  $u_y$ across a lane (red line in (c)). (h) Variation of the director angle (in radians with respect to the $y$-axis) across a lane (red line in (f)) reaching the expected Leslie angle of $ \theta_L\approx \pm 0.41$ for $\lambda=0.7$.
}
\label{fig:Laning}
\end{figure*}

\section{Emergence of laning in the flow-aligning regime}

We begin by exploring the dynamics of  flow-aligning active nematics, noting that numerical studies of the hydrodynamic screening of active nematics have  so far mostly been limited to the flow-tumbling regime~\cite{Thampi2014PRE,Doostmohammadi2016,Oza16,Pearce2019friction}. We simulate a two-dimensional active nematic with anisotropic substrate friction such that the friction in the $y$-direction $f_y$ is set to zero and the friction in the $x$-direction $f_x$ is varied, resulting in an easy flow axis along $y$. With increasing friction, the active nematic state changes from active turbulence (Fig.~\ref{fig:Laning}(a),(d) and Movies 1 and 2), to a laning state of opposing flow jets that is periodically disturbed by the nucleation of pairs of topological defects (Fig.~\ref{fig:Laning}(b),(e) and Movie 3 and 4) and then, for higher frictions, to a stable laning state without any defects (Fig.~\ref{fig:Laning}(c),(f)). 

\begin{figure*} 
    \centering
    \includegraphics[width=0.95\textwidth]{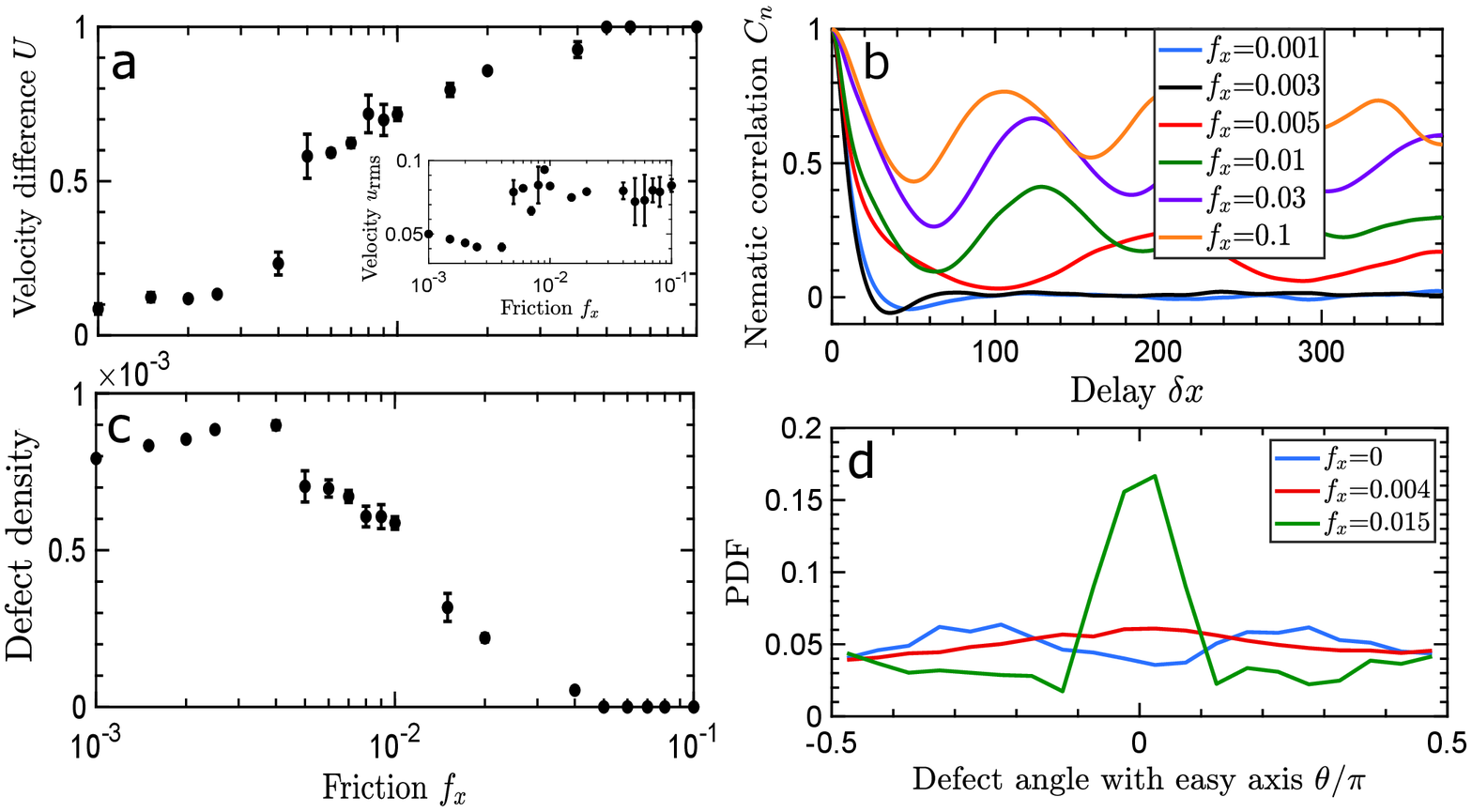}
    \caption{\textbf{Flow-aligning regime} 
(a) With increasing friction the reduced velocity difference $U = \langle (|u_y| -| u_x|)/u_\textrm{rms} \rangle$ increases from close to 0 (meaning that the velocities in $x$- and $y$- directions are approximately equal) to 1 in the state where the flow is directed solely along the $y$ axis. The inset shows the system averaged rms-velocity $u_\textrm{rms}$.
(b) Director-director correlation function for a range of frictions. Domain size for this figure was chosen as $750\times250$ lattice sites where we tripled the domain size along the anisotropic friction axis.
(c) With increasing friction, the average number of defects decreases, dropping to zero in the stable laning state.
(d) Histogram of the $+1/2$ defect orientation for $f_x=0$ (blue), $f_x=0.004$ (red) and $f_x=0.015$ (green). With increasing friction, the defects have a strong preference for moving in the positive or negative $y$-direction along the easy flow axis.
The activity is kept constant at $\zeta = 0.03$ in all graphs.
}
\label{fig:flow-aligning}
\end{figure*} 

\begin{figure*} 
    \centering
    \includegraphics[width=0.95\textwidth]{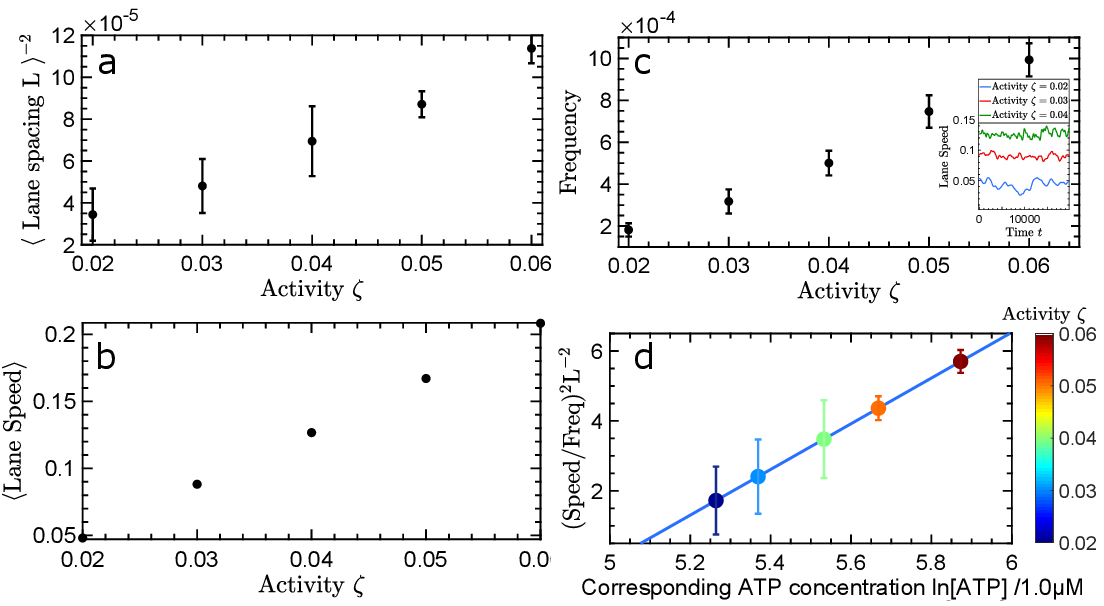}
\caption{
\textbf{Characterizing the laning state at intermediate friction} $\pmb{f_x=0.015}$\textbf{.}
(a) The distance between the lanes as a function of activity $\zeta$.
(b) The average speed of the active nematic in the lanes as a function of activity $\zeta$.
(c) The lanes still exhibit periodic instabilities due to defect creation events, resulting in fluctuations in the speed of the lanes. The mean frequency of this signal (inset) is dependent on the activity $\zeta$.
Figures (a-c) show linear dependencies with the activity corresponding to the experimentally reported relationships with the $\log{\text{ATP}}$ in~\cite{Guillamat16}.
(d) Non-dimensionalizing the numerical results (the circles) for different activities and comparing them with the experimental data (blue lines) allows us to find the ATP concentrations in a two-dimensional microtubule-kinesin mixture which correspond to the values of the active stress parameter $\zeta$ used here. Experimental data retrieved from~\cite{Guillamat16}.
}
\label{fig:Investigate_Laning}
\end{figure*} 

To further characterise this behaviour we measure the reduced velocity difference $U = \langle (|u_y| -| u_x|)/u_\textrm{rms} \rangle$ with increasing friction $f_x$ (see Fig.~\ref{fig:flow-aligning}(a)). Here $ \langle \rangle$ denotes averaging over space and time. For small friction the velocities in the $x$- and $y$-directions are approximately equal and thus $U \sim 0$. Then, as $f_x$ is increased, active flows in the $y$-direction start to dominate, leading to the emergence of the lanes along the easy flow axis. This is~marked by a sharp increase in the velocity difference $U$. Due to the ordered flow structure found in the laning state, the overall $u_\textrm{rms}$ of the system is also higher than in the state of active turbulence (Fig.~\ref{fig:flow-aligning}(a); {\it inset}).

At very high friction, the reduced velocity difference $U$ approaches $1$, corresponding to zero velocity along the $x$-axis. However, the rms velocity $u_\textrm{rms}$ remains constant in the laning state. The flow in the laning state only travels along the easy-axis, and an increase in anisotropic friction does not correspond to larger screening of the velocities. This is in stark contrast to isotropic friction and the corresponding wet to dry active nematic transition~\cite{Doostmohammadi2016}.

We also measured the director-director correlation function $C_n(\delta x) = \frac{\left\langle \bm{n}(x,y) \cdot \bm{n}(x+\delta x,y)-\frac{2}{\pi} \right\rangle_{y}}{\left\langle \bm{n}(x,y) \cdot \bm{n}(x,y)-\frac{2}{\pi} \right\rangle_{y}}$
 (Fig.~\ref{fig:flow-aligning}(b)). Because of the nematic symmetry we subtract $\frac{2}{\pi}$ to make sure that the correlation function goes to zero when there is no-long range correlation and $\langle \rangle_{y}$ denotes averaging over time and over the y-coordinate. 
The oscillations observed in this correlation function indicate that alternating ordered structures are formed, which correspond to the repeating antiparallel flow lanes. The distance between the lanes decreases with increasing friction as marked by the reduction of the oscillation wavelength at higher friction values.

The active turbulent state is characterised by topological defects in the director field, and in Fig.~\ref{fig:flow-aligning}(c) we show that the average number of defects starts to decrease at the onset of laning. A similar decrease in the defect number has been seen for isotropic friction~\cite{Doostmohammadi2016}. The defects are responsible for generating flows along $x$, and their number dropping to zero coincides with perfect laning: flows only along the $y$-axis (Fig.~\ref{fig:Laning}(f)). Moreover the $+1/2$ defects orient more strongly in the $y$-direction with increasing friction (Fig.~\ref{fig:flow-aligning}(d)) because of less resistance to their self-propulsion along $y$. Similar alignment is observed for one of the three arms of the $-1/2$ defects. The preferential alignment of motile defects to the direction of lower friction has recently been reported in \cite{Pearce2019friction}, but since that work focused on the flow-tumbling regime, no laning state was observed.

A defining feature of the laning state is diagonally-aligned nematic domains separated by straight splay-bend walls (Fig.~\ref{fig:Laning}(c),(f)). The existence of these structures is highlighted with the splay-bend order parameter $S_\text{Splay-Bend} = \partial_i \partial_j Q_{ij}$ \cite{vcopar2013}. This configuration can be explained in terms of the director response to activity-induced flows in the presence of anisotropic friction.
Stresses associated with director gradients at the splay-bend walls drive antiparallel flow lanes. Within the lanes, the director responds to the extensional part of the flow and aligns at a particular orientation with respect to the shear following the Leslie angle which is $\theta_L\approx \pm 0.41$ for $\lambda=0.7$ \cite{Voituriez2005}. We conjecture that this could allow the extraction of the flow-aligning parameter from experimental active systems. The elastic interactions favour nematic alignment, and hence a uniform shear (Fig.~\ref{fig:Laning}(g),(h)). Similar director and flow profiles are seen at the onset of flow in confined active nematics \cite{Voituriez2005, Marenduzzo2007}.

Experimentally, a laning state has been reported in two-dimensional microtubule-kinesin mixtures in contact with a passive liquid crystal~\cite{Guillamat16}.
When the passive liquid crystal is in a smectic-A phase, a magnetic field can be used to align it such that the normal to the smectic layers lies parallel to the active nematic.
The passive liquid crystal flows much more easily along the smectic layers than perpendicular to them.
Consequently, the frictional damping of the active nematic layer, due to the neighbouring smectic, is anisotropic~\cite{lubensky1996hydrodynamics} and lanes form along the easy direction, parallel to the smectic layers.
It is interesting to observe that, at least in these experiments, the active material is behaving as a flow-aligning nematic as we show in the next section that lowering the flow aligning parameter dramatically changes the behaviour of an active nematic system in the presence of anisotropic friction and destroys the laning state.

Figure 3 concentrates on an intermediate friction value $f_x=0.015$ (Fig.~\ref{fig:Laning}(b),(e)) that corresponds to a laning state, which is periodically disturbed by defect formation as seen in the experiments~\cite{Guillamat16}. At this friction the spacing between the lanes $L$ scales inversely with the square root of activity $\zeta^{-0.5}$ (Fig.~\ref{fig:Investigate_Laning}(a)), which is typically found in active hydrodynamic instabilities resulting in the formation of walls~\cite{thampi2014EPL}. Increasing the active stress also results in an increase in the overall speed of active material along the lanes. The anisotropic friction suppresses the secondary lateral instability of the walls~\cite{blow2014}, by hampering deformations and flows perpendicular to the walls. This results in stabilisation of the laminar flow as long-lived lanes. In agreement with the experimental observations~\cite{Guillamat16}, unzipping of the walls is still sometimes observed as, for this friction value, defects are continuously created and annihilated within the lanes. However, due to the suppression of the secondary instability, the walls are not destroyed. The frequency of this defect creation instability scales with the activity $\zeta$ and results in fluctuations in the speed of the lanes (Fig.~\ref{fig:Investigate_Laning}(c)), which is consistent with the experimental measurements of the fluctuations in the speed and the frequency of these fluctuations~\cite{Guillamat16}. However, in our system, as previously remarked, further increasing the friction will suppress the formation of these defects and result in the system getting stuck in a state where the spacing between lanes is sporadic.

\begin{figure*} 
    \centering
    \includegraphics[width=0.95\textwidth]{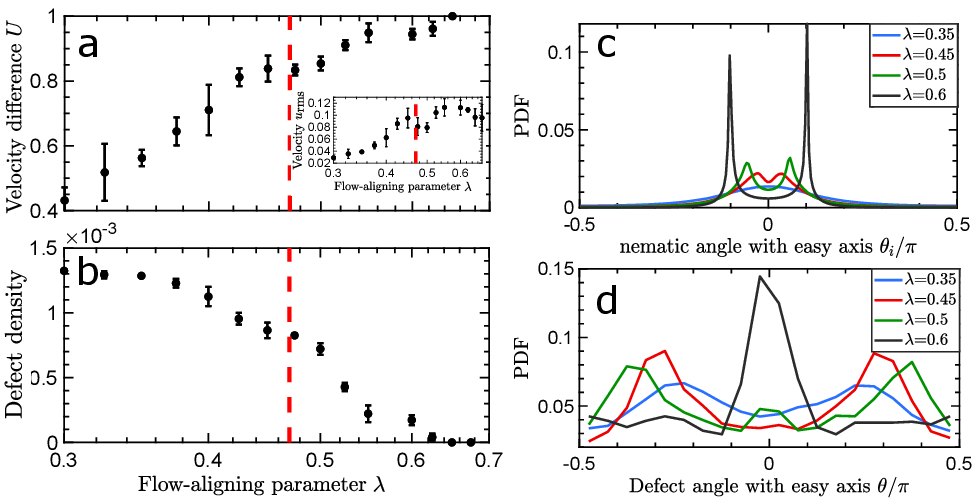}
\caption{
\textbf{Varying the flow aligning parameter.}
(a) The reduced velocity difference $U$ as a function of the flow aligning parameter with rms-velocity in the inset.
(b) Average number of defects as a function of the flow aligning parameter. The dashed red line in (a)-(b) indicates the effective flow aligning parameter $ 9q/(3q+4)\approx0.47$ transition point, corresponding to the crossover from flow-tumbling to flow-aligning behaviour.
(c) Orientation of the nematic director relative to the easy flow axis for different flow aligning parameters.
(d) Histograms of the defect orientation for different flow aligning parameters. For large flow aligning parameters defects move mostly along $y$, but for smaller flow aligning parameters additional peaks appear around $\theta \sim \pm0.2\pi$.
}
\label{fig:vary_tumbling}
\end{figure*} 

In addition to observing a similar qualitative behaviour as in the experiments, the quantitative measurements of the spacing between the lanes, the velocities within the lanes, and the frequency of the periodic instabilities within the lanes, allows us to map the activity coefficient $\zeta$ to the ATP concentrations used in the experiments. To this end, we define a dimensionless parameter based on the measurements of the speed, frequency and the length scale associated with the spacing between the lanes: $\Lambda = \left({\text {Speed}}/{\text {Freq}}\right) ^2/ L^2$. It is expected that the activity coefficient is proportional to the chemical potential of ATP hydrolysis $\zeta\propto {\text{ln(ATP)}}$. As such comparing the dimensionless number $\Lambda$ extracted from the experiments as a function of ${\text{ln(ATP)}}$ with that from the simulations for different values of activity $\zeta$ (Fig.~\ref{fig:Investigate_Laning}(d)) results in a direct mapping of the activity coefficient to the ATP concentrations used in the experiments~\cite{Guillamat16}.

\section{The flow tumbling regime}

The crossover from flow-aligning to flow-tumbling should occur at an effective flow aligning parameter  $9q/(3q+4)\approx0.47$~\cite{BerisBook}. This definition depends on the value of the nematic order $q$. For high activity systems, $q$ increases
from its equilibrium value~\cite{thampi2015}. As a result, the transition from flow-aligning to flow-tumbling, does not occur exactly at the calculated equilibrium parameter in Fig. \ref{fig:vary_tumbling}(a,b), but at a slightly higher flow-aligning value. In the flow-tumbling regime, the director rotates in a shear flow and therefore the laning state is expected to be unstable. To investigate this, we fix the friction to $f_x=0.1$ and the activity to $\zeta=0.06$ (well within the laning regime), while we reduce the flow aligning parameter. As evident in Fig.~\ref{fig:vary_tumbling}, both the velocity difference $U$ and the rms-velocity $u_\text{rms}$ decrease as the flow aligning parameter is reduced (Fig. \ref{fig:vary_tumbling}(a)), while the number of topological defects increases from its zero value in the laning state (Fig. \ref{fig:vary_tumbling}(b)). The director moves away from the Leslie angle to a more isotropic distribution, but with the expected preference for the easy flow axis (Fig. \ref{fig:vary_tumbling}(c)).

\begin{figure*} 
    \centering
    \includegraphics[width=0.95\textwidth]{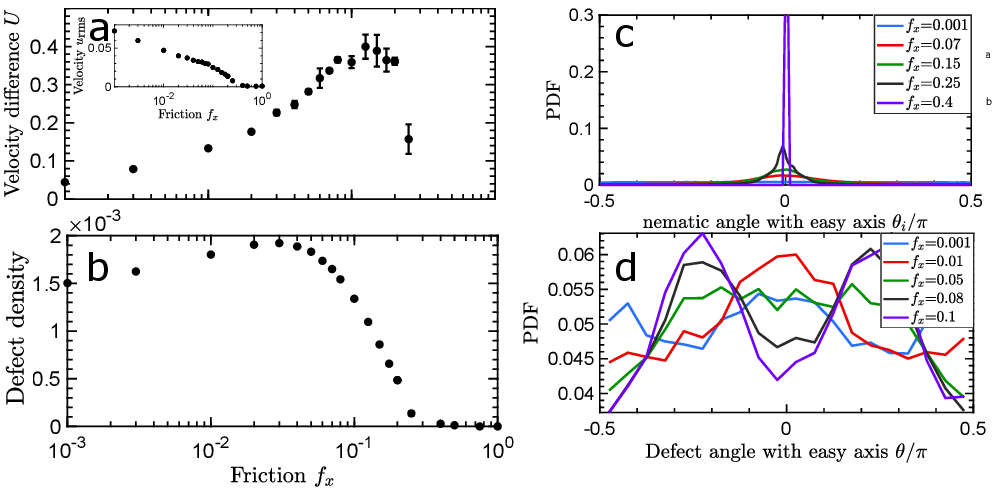}
\caption{
\textbf{Flow-tumbling regime.}
(a) The reduced velocity difference $U$ as a function of the friction with rms-velocity in the inset. Velocity difference $U$ increases with increasing friction until the friction becomes so strong that any flow is suppressed in the system, and the reduced velocity difference becomes ill-defined.
(b) Average number of defects as a function of the friction.
(c) Orientation of the nematic director relative to the easy flow axis for different frictions.
(d) Histograms of the $+1/2$ defect angle for different frictions. At large friction, the defects start moving preferentially at a given angle with respect to the easy flow axis, which is illustrated by peaks forming at different positions.
}
    \label{fig:tumbling}
\end{figure*} 

An interesting change in the behaviour of the system with decreasing flow alignment parameter is observed in the angular distribution of the motile $+1/2$ defects (Fig. \ref{fig:vary_tumbling}(d)). We showed that for large flow aligning parameters, where the system is in the flow-aligning regime and the laning state is established, the defects are mostly oriented along the easy flow axis. This is marked by a single peak in the defect angle distribution (Fig.~\ref{fig:vary_tumbling}(d); {\it black solid line}). However, as the flow aligning parameter is reduced, two additional peaks appear in the histogram. These correspond to defects moving at some angle with respect to the $y$-axis. At yet smaller flow aligning parameters, where the reorientation of the director is predominantly controlled by the rotational part of the flow gradient, any preference for moving along the $y$-axis disappears, and only two peaks remain (Fig.~\ref{fig:vary_tumbling}(d); {\it red, blue solid lines}). This occurs when flows in $x$-direction are dissipated quickly, strongly suppressing self-motility of $+1/2$ defects along $x$. At the same time, the flow-tumbling behaviour of the active nematic means that the $+1/2$ defect aligned along $y$ wants to rotate. These two counterbalancing effects lead to a state where defects travel at a preferred angle with respect to the $y$-axis.

In order to further check the behaviour of the active material in the flow-tumbling regime, we next fix the flow aligning parameter to $\lambda=0.3$, the activity to $\zeta=0.06$, and vary the friction. The rms velocity steadily decreases with the increase in friction (Fig.~\ref{fig:tumbling}(a)) and, after an initial increase, the number of topological defects falls to zero (Fig.~\ref{fig:tumbling}(b)). The reduced velocity difference $U$ increases initially, but then starts to drop down and eventually becomes ill-defined as the rms velocity goes to zero. The director field gradually develops a preference for the easy flow direction until, for sufficiently strong friction, tumbling is suppressed and nematic alignment is imposed by the flow (Fig.~\ref{fig:tumbling}(c)). Fig.~\ref{fig:tumbling}(d) shows the orientation of the defects as a function of friction. Again for higher frictions the two-peak structure is clearly visible. For lower frictions the self-propulsion defect velocity dominates and they tend to move along the easy flow axis.

\begin{figure*} 
    \centering
    \includegraphics[width=0.95\textwidth]{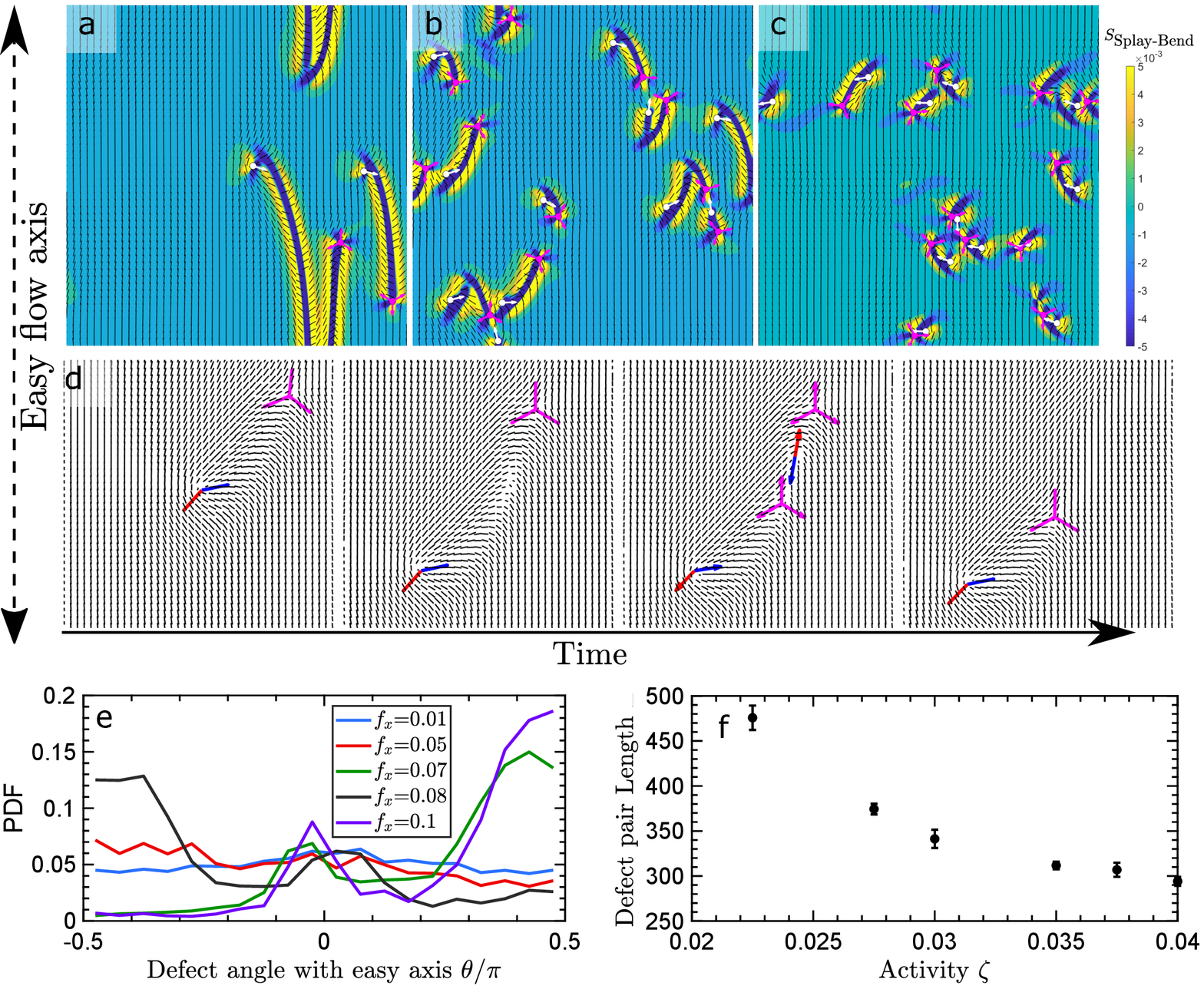}
\caption{
\textbf{Bound defect pairs at low activities.}
(a)-(c) Snapshots of different bound defect states at different activities: (a) $\zeta=0.0175$, (b) $\zeta=0.0275$ and (c) $\zeta=0.0375$. Friction is $f_x=0.3$.  White lines denote the orientation of $+1/2$ defects and purple trefoils denote the $-1/2$ defects (see also Movies 5-7).
(d) Simulation results showing the propagation of a defect pair with time. Red (blue) arrows show the $+1/2$ defect velocity direction (orientation). Purple trefoils denote the $-1/2$ defects. Simulation values are $\zeta=0.04$ and $f_x=0.5$. Snapshots are taken at times 0, 1700, 1800 and 2000 (see also Movie 8).
(e) Histogram of the defect angles for different frictions. For higher frictions the system spontaneously breaks the symmetry with the scars in the nematic ordering between the defects preferring either a clockwise or an anticlockwise angle with respect to the easy flow axis.
(f) The length of scars. This is calculated by calculating the area where the bend-splay order parameter $S_{Splay-Bend}<-0.01$ and then dividing by the number of defect pairs.
\label{fig:scarring}}
\end{figure*} 

\begin{figure*}
    \centering
    \includegraphics[width=0.85\textwidth]{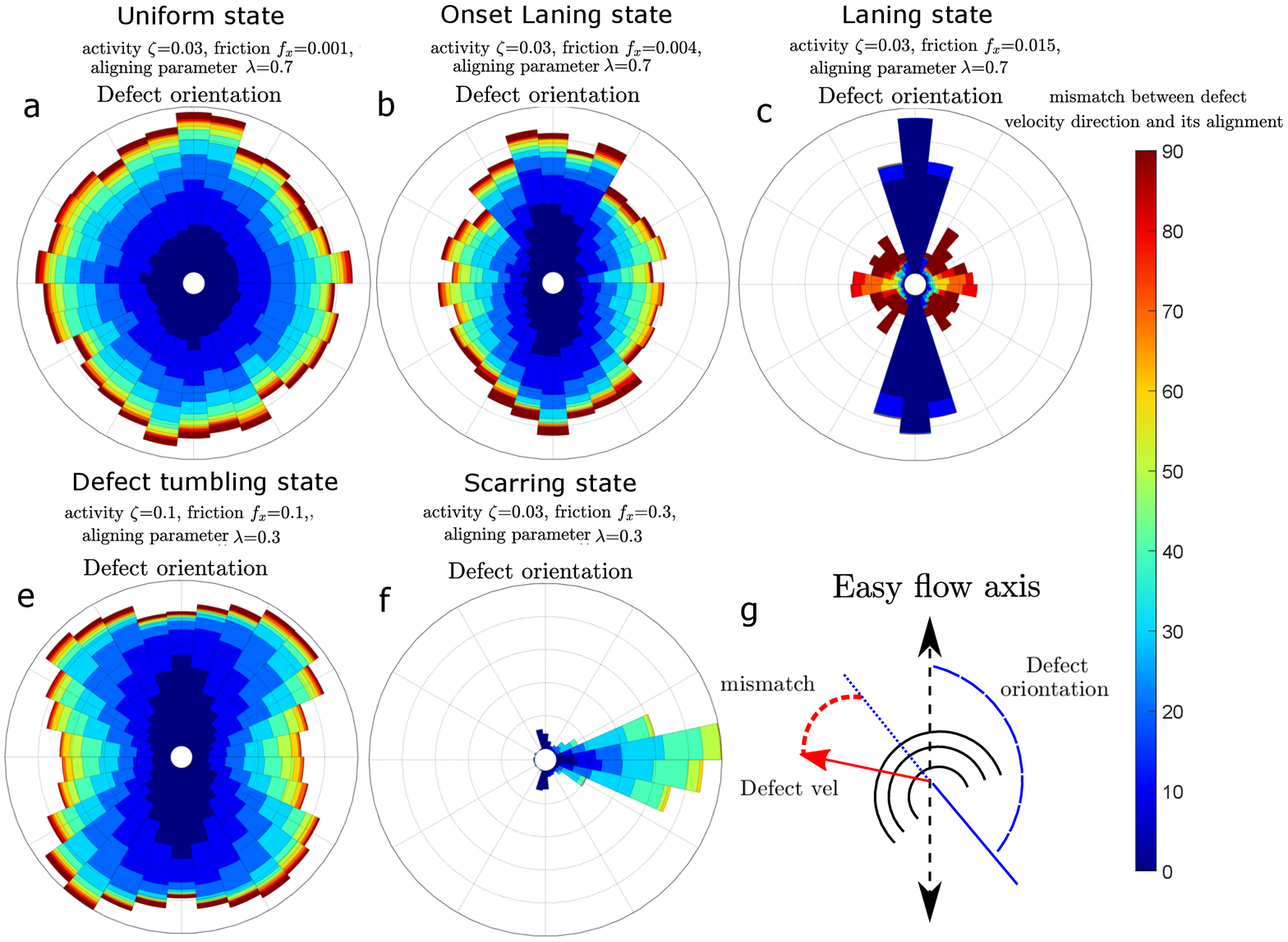}
\caption{\textbf{Wind roses of the different states.}
The $+1/2$ defect orientation and its mismatch with the velocity orientation at the defect core (color map)  for (a) small friction anisotropy,
(b) just at the onset of the laning state,
(c) for the stable laning state,
(d) within the tumbling regime, and
(e) for scarring state.
(f) Schematic showing the quantities of interest for the measurements in (a)-(e).     \label{fig:roses}}
\end{figure*} 

\section{A state of bound defect pairs}

If the activity is lowered further within the tumbling regime a new feature appears in the defect dynamics (Fig.~\ref{fig:scarring}(a)-(c) and Movies 5-7). There is a breaking of the mirror symmetry and only one preferred angle is observed (Fig.~\ref{fig:scarring}(e); {\it green, black and purple lines}).
In order to better understand how this mirror symmetry-breaking mechanism works, we next focus on the high friction regime and very small activities such that only a few defects are created within the director field and their dynamics can be closely monitored (Fig.~\ref{fig:scarring}(d) and Movie 8). Upon nucleation of defect pairs - along the direction with lower friction ($y$-direction) - the defects initially move away from each other, trying to unbind. However, the activity is too weak to overcome the elastic attraction and the defect pair remains bound. At the same time, since the system is in the flow-tumbling regime, the defects slightly rotate relative to each other, so that the line of director distortions in between the defects is no longer along the $y$-direction. The bound defect pair starts moving as a single object at an angle with respect to the $y$-direction, with the $+1/2$ defect dragging the $-1/2$ defect behind. The director distortions between the defects appear as `scars' on the background nematic ordering (Fig.~\ref{fig:scarring}(a)-(c)). 

With time an additional pair can be nucleated at a scar, with the new $+1/2$ defect annihilating the original $-1/2$ defect, and the new $-1/2$ defect binding to the original $+1/2$ defect. Increasing activity leads to a larger number of defect nucleation events and shorter distances between the defects in a bound pair (Fig.~\ref{fig:scarring}((a)-(c),(f)). The scarred defects are descendants of an original defect pair, and due to the frictional damping, they retain the original orientation after creation. This results in the scars tending to point in the same direction over long time scales. The scar dynamics is reminiscent of the polar ordering of $+1/2$ topological defects that was observed in a study of over-damped active nematics with broken Galilean invariance~\cite{putzig16}.

\section{Discussion}

The results presented in this paper demonstrate the important interplay between the flow-aligning behaviour of active particles and anisotropy in hydrodynamic screening. By varying the flow aligning parameter from flow-aligning to flow-tumbling, we are able to consolidate different observations of active nematic behaviour in anisotropic environments including the laning state in the flow-aligning regime and preferential alignment of $+1/2$ defects in the flow-tumbling regime. Moreover, when the nematic is flow-tumbling, and at low activities, we identify long-lived and motile bound defect pairs that spawn new pairs and that appear as scar-like distortions within the otherwise aligned nematic.

In order to more clearly summarise and distinguish the different behaviours observed, we generate wind rose plots of $+1/2$ defect alignment with respect to the direction with lower friction and coloured by the angular mismatch between the alignment of the $+1/2$ defect and the direction in which it moves (Fig.~\ref{fig:roses}). In the absence of any friction anisotropy, the angular distribution would be uniform and the $+1/2$ defects self-propel along their direction of alignment (mismatch angle will be identical to zero) regardless of the flow aligning parameter.

Let us begin with the flow-aligning case and increasing friction: at low friction (Fig.~\ref{fig:roses}(a)) the angular distribution of $+1/2$ defects is uniform as expected, but due to the active turbulent flows there is the possibility of mismatch between a defect's orientation and its velocity. This starts to change as the friction anisotropy is increased, such that at the onset of lane formation (Fig.~\ref{fig:roses}(b)), defects show a higher tendency to be aligned along the easy flow axis, and they predominantly move along this direction. Deep into the laning state (Fig.~\ref{fig:roses}(c)) the majority of $+1/2$ defects align along the easy flow axis, and any mismatch between their orientation and their velocity is greatly diminished. There is also a small number of defects pointing perpendicular to the easy flow axis. These are defects just after nucleation events, where wall formation results in their orientation being perpendicular to the easy flow axis. However, due to the large velocity gradients in between the lanes, these defects rapidly turn to align with the easy flow axis.

A different behaviour of the active defects is observed in the flow-tumbling regime ({\it bottom row in} Fig.~\ref{fig:roses}): At low friction anisotropy and high activity the $+1/2$ defects predominantly align and move along the easy flow axis and two other preferred directions (Fig.~\ref{fig:roses}(d)). This is consistent with the behaviour reported in \cite{Pearce2019friction} and is due to the balance of the friction anisotropy trying to reorient defects along the easy flow axis with the flow-tumbling behaviour. Increasing the friction anisotropy and reducing activity dramatically changes this picture (Fig.~\ref{fig:roses}(e)): The $+1/2$ defects align predominantly away from the easy flow axis, but move at an angle that has a large mismatch with their alignment, creating the scarring state introduced in the previous section. 

It is noteworthy that we do not observe a similar laning behaviour for contractile active nematics as the splay deformations that form at the centre of the lanes are unstable for contractile stresses. On the other hand, the possible reverse case of the stabilization of bend at the centre of the lanes does not work since this would result in the situation that the nematic director and the velocity orientation are perpendicularly orientated, which is not stable due to the flow-aligning effect. Therefore, in the contractile case, we retain the active chaotic flows with increasing anisotropic friction. The defects become more aligned, similar to the results found in~\cite{Pearce2019friction}, until the flows are completely killed due to very high frictional damping.

\section{Conclusions}
We have numerically investigated the impact of anisotropic hydrodynamic screening on the behaviour of active nematics by imposing anisotropic friction between an active nematic and its surroundings. We find that competing effects between the torques induced by frictional anisotropy and the flow-induced torques result in dynamics that are strongly sensitive to the flow aligning parameter that characterises the aspect ratio of elongated particles. At high values of the flow aligning parameter, where the orientation of particles is predominantly affected by the extensional flows, we recover and explain the laning state previously reported in microtubule-motor protein mixtures in contact with a smectic liquid crystal~\cite{Guillamat16}. From this, we have shown that it would be possible to extract the flow aligning parameter and get an estimate of the corresponding active stresses from experimental data. As the alignment parameter is reduced to values corresponding to the flow tumbling regime, in which rotational flows predominantly determine particle orientations, we observe a dramatic change in the flow patterns. The laning state disappears and the self-propelled $+1/2$ defects align along specific orientations relative to the direction of the friction anisotropy. Furthermore, at low flow aligning parameter values and small activities, we find a peculiar arrangement of topological defect pairs, where bound $\pm 1/2$ topological defects navigate through the otherwise aligned active nematic, breaking the mirror-symmetry of the system and creating scar-like distortions in the director field. In addition to confirming the mechanism of previous experimental observations, our work could motivate new sets of experiments exploring the effects of the flow aligning parameter on the patterns of motion in active nematic materials as our results showcase that the flow aligningparameter results in distinct behaviours associated with flow-tumbling or flow-aligning behaviour.


\section*{Acknowledgements}
This project has received funding from the European Union's Horizon 2020 research and innovation programme under the Lubiss the Marie Sklodowska-Curie Grant Agreement No. 722497 for K.T. and DiStruc Marie Sklodowska-Curie Grant Agreement No. 641839 for L.M. Additionally, A.D acknowledges support from the Novo Nordisk Foundation (grant agreement No. NNF18SA0035142).
\section{References}

\bibliographystyle{apsrev4-1}
\bibliography{Ani_friction}

\section{Movie captions}

\textbf{Movie 1:}  Movie corresponding to Fig. 1(a), showing the flow field in active turbulence. Velocity field is coloured by the magnitude of the rms-velocity $u_\text{rms}$. White circles denote $+1/2$ defects with the attached lines indicating their orientation pointing from head-to-tail and purple trefoils denote $-1/2$ defects. Time between every frame is 50 simulation timesteps. 
\par
\textbf{Movie 2:} Movie corresponding to Fig. 1(d), showing the director field in active turbulence. Director field is coloured by the magnitude of the splay-bend order parameter $S_\text{Splay-Bend} = \partial_i \partial_j Q_{ij}$. White circles denote $+1/2$ defects with the attached lines indicating their orientation pointing from head-to-tail and purple trefoils denote $-1/2$ defects. Time between every frame is 50 simulation timesteps.
\par
\textbf{Movie 3:} Movie corresponding to Fig. 1(b), showing the flow field in the laning state. Velocity field is coloured by the magnitude of the rms-velocity $u_\text{rms}$. White circles denote $+1/2$ defects with the attached lines indicating their orientation pointing from head-to-tail and purple trefoils denote $-1/2$ defects. Time between every frame is 50 simulation timesteps.
\par
\textbf{Movie 4:} Movie corresponding to Fig. 1(e), showing the director field in the laning state.  Director field is coloured by the magnitude of the splay-bend order parameter $S_\text{Splay-Bend} = \partial_i \partial_j Q_{ij}$. White circles denote $+1/2$ defects with the attached lines indicating their orientation pointing from head-to-tail and purple trefoils denote $-1/2$ defects. Time between every frame is 50 simulation timesteps.
\par
\textbf{Movie 5:} Movie corresponding to Fig. 5(a), showing the director field in the scarring state for low activity. Defect pairs are bound, but the $-1/2$ defect is not dragged along with the $+1/2$ defect resulting in long scars. Director field is coloured by the magnitude of the splay-bend order parameter $S_\text{Splay-Bend} = \partial_i \partial_j Q_{ij}$. White circles denote $+1/2$ defects with the attached lines indicating their orientation pointing from head-to-tail and purple trefoils denote $-1/2$ defects. Time between every frame is 1000 simulation timesteps.
\par
\textbf{Movie 6:} Movie corresponding to Fig. 5(b), showing the director field in the scarring state for medium activity.  Defect pairs remain tightly bound. However, the $-1/2$ defect remains some distance from the $+1/2$ defect.   Director field is coloured by the magnitude of the splay-bend order parameter $S_\text{Splay-Bend} = \partial_i \partial_j Q_{ij}$. White circles denote $+1/2$ defects with the attached lines indicating their orientation pointing from head-to-tail and purple trefoils denote $-1/2$ defects. Time between every frame is 1000 simulation timesteps.
\par
\textbf{Movie 7:} Movie corresponding to Fig. 5(c), showing the director field in the scarring state for high activity. Defect pairs remain tightly bound, and many pairs are formed.  Director field is coloured by the magnitude of the splay-bend order parameter $S_\text{Splay-Bend} = \partial_i \partial_j Q_{ij}$. White circles denote $+1/2$ defects with the attached lines indicating their orientation pointing from head-to-tail and purple trefoils denote $-1/2$ defects. Time between every frame is 1000 simulation timesteps.
\par
\textbf{Movie 8:} Movie corresponding to Fig. 5(d), showing propagation of a defect pair and the creation of a new one. Red (blue) arrows show the $+1/2$ defect velocity direction (orientation). Purple trefoils denote the $-1/2$ defects. Black lines show the director field. Time between every frame is 10 simulation timesteps. 

\end{document}